\documentclass[11pt,twoside]{article}

\usepackage{asp2006}
\usepackage{epsf}
\usepackage{psfig}
\usepackage{lscape}

\begin{document}
\title{Introducing the H-Index in Telescope Statistics}   
\author{Uta Grothkopf}   
\affil{European Southern Observatory, Karl-Schwarzschild-Str. 2, 85748
Garching, Germany}    

\author{Sarah Stevens-Rayburn}   
\affil{Space Telescope Science Institute, 3700 San Martin Drive,
Baltimore, MD 21218, USA}    

\begin{abstract}
This paper analyzes the performance of observatories based on the
so-called {\itshape h}-index \citet*{hirsch2005}, a new, easy-to-use
parameter that quantifies  scientists' research impact and
relevance. Compared to  other bibliometric criteria, like total number
of publications  or citations, the {\itshape h}-index is less biased. Using
NASA's Astrophysics Data  System (ADS), we investigate the performance
of selected  observatories, taking into account their specific number
of years of  operation. 
\end{abstract}


\section{Telescope Statistics}
At many observatories, librarians are engaged
in compiling  telescope statistics. These compilations of papers that
use data coming  from specific observing facilities are needed for a
number of purposes,  among them assessing the overall impact of
facilities, measuring  the scientific output from observing programs,
and reporting back  to funding authorities. Despite the intrinsic
dangers (see below),  telescope statistics are also often used to
compare output among observatories.  

Certain bibliometric methods are used frequently in order to evaluate
facilities.  These methods usually show some advantages as well as specific
disadvantages. For instance, counting the total
number of  publications measures productivity of an observatory, but
does not take  the importance and impact into account which the papers
resulting  from the data have generated. In contrast, looking at the
total number  of citations does measure impact, but the value may be
inflated by  a few papers that received extraordinarily high numbers
of citations.  Another method is to investigate contributions to the
so-called High-Impact Papers,  a term originally coined by Thomson
ISI for their database of the most influential papers in specific
fields.  In astronomy, this name is
often used to  refer to a study performed by the ST ScI, where
the 200 highest  cited observational papers are
distributed across  observatories, based on their actual input of data
to these papers (see Madrid, these proceedings).
Measuring contributions 
to High-Impact  Papers is more balanced than other bibliometric
methods, but the  values per observatory are much more difficult to find. 

\section{The {\itshape h}-Index}
In 2005, Jorge E. Hirsch of the University of California at San
Diego introduced  a new measure, the so-called {\itshape h}-index
\citet*{hirsch2005}.  The original  aim was to quantify an
individual's scientific 
research output.  According to the definition by Hirsch, a researcher
with index {\itshape h}  has {\itshape h} papers with at least
{\itshape h} citations.  In other words,
index {\itshape h} is ``the highest number of papers a scientist has that have
each received  at least that number of citations'' \citet*{ball2005}. For
instance, if a  researcher has written 50 papers, 30 of which have achieved
30 or more citations,  his or her {\itshape h}-index is 30. 

In order to find the {\itshape h}-index, one needs a list of all papers
fulfilling the criteria  under investigation (e.g., written by a
specific person  or using data from a given facility). This list must
be numbered (from 1 to the total number of papers in the list) and
ranked by decreasing  citation counts, i.e., the paper with the
highest number of  citations appears on top, that with the lowest number at
the bottom. {\itshape h} can be found where  the number of citations
are at least as high as the rank (publication number). See Fig. 1 for
an example,  using the ADS. Fig. 2 demonstrates that neither a large number of 
papers with few or no cites (x-axis, right end) nor individual papers
with a very high number of citations (y-axis, upper end) influence {\itshape h}
considerably. 
\vspace{1.5cm}

\begin{figure}[!ht]
\plotone{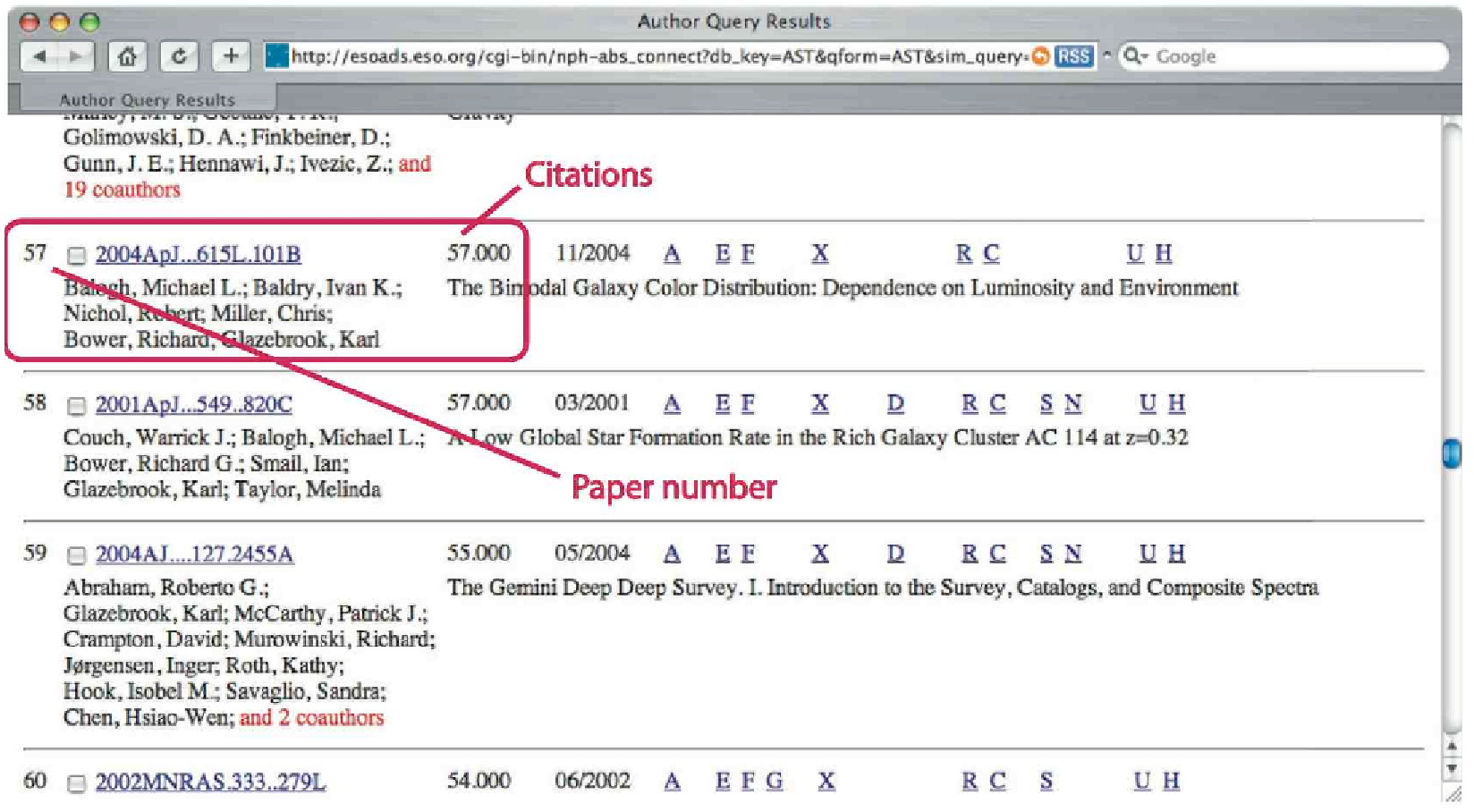}
\caption{How to find the {\itshape h}-index with the ADS: the
highest paper number that has at least as many citations.}
\end{figure}

\begin{figure}[!ht]
\epsfxsize=2cm
\centerline{\epsfxsize=9cm \epsffile{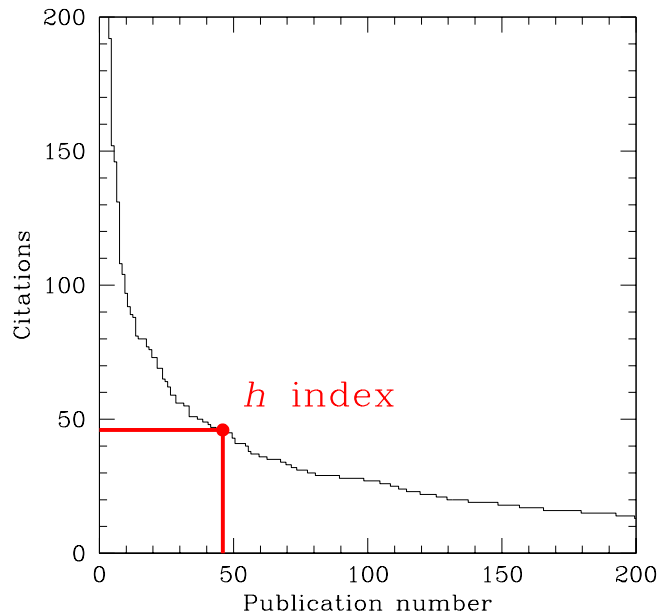}}
\caption{{\itshape h}-index: where the rank (counted publication
number) and the number of citations are the same.}
\end{figure}

\section{Applying {\itshape h} to Observatories}
The authors assumed that what works well for individual researchers
should also be  applicable to facilities, and applied Hirsch's
{\itshape h}-index to observatories. Our study includes papers from
the 
CFHT, HST, Keck, ESO VLT and Gemini.  This selection is somewhat
arbitrary as it  was largely defined by the question of which observatories
provide easy  access to bibcodes of papers resulting from their data. 
 
Before we continue, we would like to issue a word of warning: it is
very difficult,  if not impossible, to compare statistics across
observatories in  a fair, unbiased way. Some of the reasons originate
from the procedures -- the
criteria for inclusion  of papers are not uniform so that the
contents of the resulting bibliographies vary. Also the
methodologies applied for building telescope 
bibliographies vary  among institutions; they can range from 
electronic searching whatever sections of relevant
journals are available online to
scrutinizing journal issues on paper (with the non-exclusively
electronic approach still leading to a higher completeness, see
Stevens-Rayburn \& Grothkopf, these proceedings). Other difficulties
result from the differences among facilities: we only have to think of
ground-based versus space-based  observatories, different
apertures, different wavelengths etc. Therefore,  our study is meant
as a first introduction of the {\itshape h}-index in  the evaluation
of facilities; further studies will be needed to fine-tune the methodology.

\section{Our Sample}
As mentioned above, our sample includes those telescope bibliographies
to which we  had comparably easy access. These are the following
observatories and  ranges of years of coverage:

\begin{table}[!ht]
\caption{Total sample included in our study.}
\smallskip
\smallskip
\begin{center}
{\small
\begin{tabular}{lc}
\noalign{\smallskip}
\tableline
\noalign{\smallskip}
CFHT & 1980 - 2005\\ 
\noalign{\smallskip}
HST & 1991 - 2005\\
\noalign{\smallskip}
Keck & 1996 - 2005\\
\noalign{\smallskip}
ESO VLT & 1999 - 2005\\
\noalign{\smallskip}
Gemini & 2000 - 2005\\
\noalign{\smallskip}
\tableline
\end{tabular}
}
\end{center}
\end{table}

For uniformity, we ended the sample with publication year 2005 for all
facilities.  

For the years given above, we collected all bibcodes of papers
included in the observatories' telescope bibliographies as follows:
for CFHT  and HST, bibcodes were retrieved through the
ADS by using the filter ``Select References In:'' CFHT and HST,
respectively, sorted by citation count. (This filter also allows to
select ESO Telescopes; however, this feature does not discriminate
between ESO VLT and ESO La Silla papers. Such a distinction is only
possible through the ESO Telescope Bibliography, see below.) Keck
bibcodes were extracted 
from the references provided in their Science Bibliography
(http://www2.keck.hawaii.edu/library/1996.htm through
2005.htm). Gemini bibcodes and numbers of citations were
kindly provided by the Gemini librarian. Bibcodes of papers using ESO VLT
data are available through the ESO Telescope Bibliography
(http://www.eso.org/libraries/telbib.html).

Individual citations per paper were retrieved from the ADS and (in the
case of Keck, VLT and Gemini) listed in decreasing order in an Excel
spreadsheet so that the {\itshape h}-index could be obtained. For CFHT
and HST, the ADS very conveniently provides the {\itshape h}-index as
described above. 

Computing the {\itshape h}-index for all years included in our study
led to the following results: CFHT (1980 - 2005): 94; HST (1991 -
2005): 131; Keck (1996 - 2005): 105; ESO VLT (1999 - 2005): 69; Gemini
(2000 - 2005): 26. Note that in order to be meaningful, {\itshape h}
needs to be 
complemented with a parameter that normalizes based on the number of
years of operation!

\section{{\itshape h}-Index and {\itshape m} Parameter }
Obviously, the telescopes we investigated did not come online at the
same time, but over a period of 20 years (see Table 1). Accordingly,
some had more, others much less time to gather citations. We therefore
applied a measure to normalize results based on the time that has
lapsed since the first papers were published: the {\itshape m}
parameter, also introduced by Hirsch (2005). In simple words,
{\itshape m} is {\itshape h} divided by the number of years of
operation. For instance, if {\itshape h} is 20 after 20 years,
{\itshape m} is 1 (20 : 20 = 1); if {\itshape h} is 40 after 20 years,
{\itshape m} is 2 (40 : 20 = 2) etc. The higher {\itshape m} is, the
better.  

In order to retrieve {\itshape m} easily for our sample, we introduced
a ``Years since first paper'' column in our table and divided {\itshape
h} by the number of years indicated in this column. The results are
shown in Table 2 and Fig. 3.

\begin{table}[!ht]
\caption{{\itshape h}-index and parameter {\itshape m} computed for
all publication years included in our study.}
\smallskip
\smallskip
\begin{center}
{\small
\begin{tabular}{lcccc}
\tableline
\noalign{\smallskip}
Observatory & Years & Years since first paper & {\itshape h} &  {\itshape m}\\
\noalign{\smallskip}
\tableline
\noalign{\smallskip}
CFHT & 1980 - 2005 & 25 & 94 & 3.76\\ 
\noalign{\smallskip}
HST & 1991 - 2005 & 14 & 131 & 9.36\\
\noalign{\smallskip}
Keck & 1996 - 2005 & 9 & 105 & 11.67\\
\noalign{\smallskip}
ESO VLT & 1999 - 2005 & 6 & 69 & 11.5\\
\noalign{\smallskip}
Gemini & 2000 - 2005 & 5 & 26 & 5.2\\
\noalign{\smallskip}
\tableline
\end{tabular}
}
\end{center}
\end{table}
\vspace{-1.33cm}

\begin{figure}[!ht]
\plotone{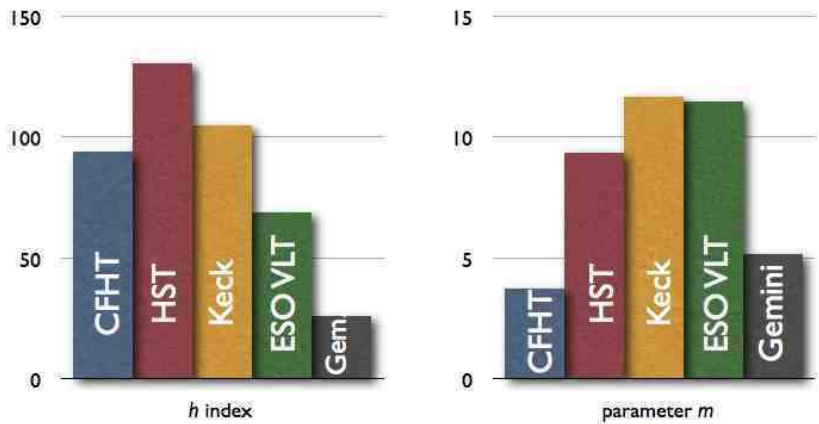}
\caption{{\itshape h}-index and parameter {\itshape m} of our entire
sample (see Table 1). Note that {\itshape h}
needs to be complemented with {\itshape m} to be meaningful.}
\end{figure}

\section{{\itshape h} and {\itshape m} for Identical Range of Years}
In order to focus on common years of operation, we went on to
investigate those telescopes for which we had bibliographies stemming
from the same period, namely CFHT, HST and Keck. For these facilities,
we limited the range of years to 1996 - 2005 and again computed
{\itshape h} and {\itshape m}. The results are given in Table 3. While {\itshape h} and {\itshape m} were vastly different when
looking at all years, the values take on a much more similar shape for
the subset of years (even though on a different scale, see Fig. 4),
with HST being the leader, followed by 
Keck and CFHT. In comparison to results of our entire study (from the
first papers onwards), {\itshape m}-values of the selected years 1996
to 2005 have increased for the two older telescopes (CFHT and HST),
suggesting that the metric doesn't flow as linearly as one might
desire, but rather new instrumentation or more creative observing
proposals can positively affect the citation rates for older observatories.

\begin{table}[!ht]
\caption{{\itshape h}-index and parameter {\itshape m} computed for
the years 1996 - 2005.}
\smallskip
\smallskip
\begin{center}
{\small
\begin{tabular}{lcccc}
\tableline
\noalign{\smallskip}
Observatory & Years & Years since first paper & {\itshape h} &  {\itshape m}\\
\noalign{\smallskip}
\tableline
\noalign{\smallskip}
CFHT & 1996 - 2005 & 9 & 62 & 6.89\\ 
\noalign{\smallskip}
HST & 1996 - 2005 & 9 & 118 & 13.11\\
\noalign{\smallskip}
Keck & 1996 - 2005 & 9 & 105 & 11.67\\
\noalign{\smallskip}
\tableline
\end{tabular}
}
\end{center}
\end{table}
\vspace{-1.2cm}

\begin{figure}[!ht]
\plotone{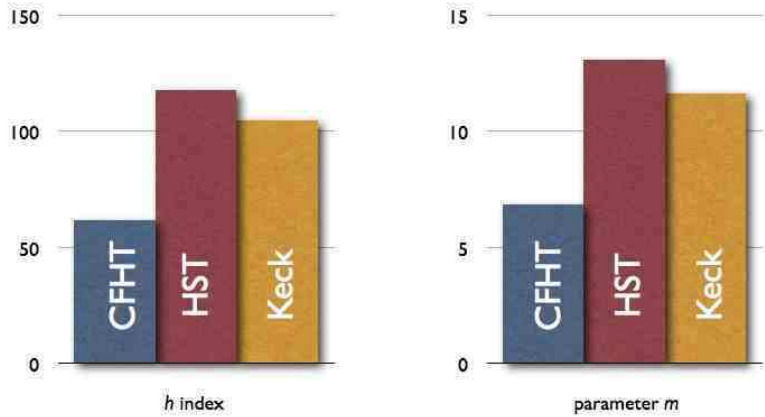}
\caption{{\itshape h}-index and parameter {\itshape m} of selected
telescope bibliographies for the years 1996 - 2005.}
\end{figure}

For the HST, we conducted another brief study in order to find out if
and how {\itshape h} and {\itshape m} change over time. For all HST papers,
going back to the first ones published after the launch in 1991, and including
the first six months of 2006, we found {\itshape h} = 132 and
{\itshape m} = 8.8. If the early papers are left out and the range of
years taken into account starts after the first servicing mission
(COSTAR) in 1994, these values change to {\itshape h} = 128 and
{\itshape m} = 10.66. STIS and NICMOS were installed in 1998, and
{\itshape h} and {\itshape m} for the years 1998 to mid-2006 are 103
and 12.87, respectively. This indicates that also the {\itshape m}
parameter can be misleading with regard to telescopes that continue to
be refurbished over time. Both {\itshape h} and {\itshape m} therefore
should be used in telescope statistics with greatest care.

\section{Conclusions}
There are some clear advantages of the {\itshape h}-index which may
make it suitable also for telescope statistics:

\begin{itemize}
\item it combines productivity and impact
\vspace{-.2cm}

\item it is easy to find using the ADS
\vspace{-.2cm}

\item it is not sensitive to extreme values
\end{itemize}

On the other hand, as with all statistics, one has to apply utmost
attention when applying the {\itshape h}-index, in particular because

\begin{itemize}
\item it is determined by the number of years of operation and
\vspace{-.2cm}

\item it needs to be combined with the {\itshape m} parameter for
comparison across facilities
\end{itemize}

In this paper,
we have shown how the {\itshape h}-index and the {\itshape m}
parameter can be used in telescope statistics. However,  it is important to note
that one single parameter is never enough to represent the complex reality,
and a narrowly based approach might
lead to incorrect  results and wrong interpretations. 
\vspace{1cm}


\acknowledgements
The authors wish to thank Peggy Kamisato (Keck Observatory), Liz
Bryson (CFHT) and Xiaoyu  Zhang (Gemini Observatory) for providing
access to their  telescope bibliographies, either personally or
through the ADS. We are very grateful to Juan Madrid (ST ScI), Don
Stevens-Rayburn and Sandra Savaglio (MPE) for extracting and comparing
Keck bibcodes. Special thanks go to Elizabeth
Fraser (ST ScI) and Angelika Treumann (ESO) for maintaining the ST ScI
and ESO telescope bibliographies.


\end{document}